# Vertical One-Dimensional Photonic Crystal Platforms for Label-Free (Bio)Sensing: Towards Drop-And-Measure Applications

Giuseppe Barillaro
Dipartimento di Ingegneria dell'Informazione,
Università di Pisa, via G: Caruso 16, 56122 Pisa, Italy
g.barillaro@iet.unipi.it

*Abstract*— **In this work, all-silicon, integrated optofluidic platforms, fabricated by electrochemical micromachining technology, making use of vertical, one-dimensional high-aspect-ratio photonic crystals for flow-through (bio)sensing applications are reviewed. The potential of such platforms for point-of-care applications is discussed for both pressure-driven and capillarity-driven operations with reference to refractometry and biochemical sensing.**

## I. INTRODUCTION

In the last two decades, one-dimensional (1D) photonic crystals (PhCs), as well as 2D and 3D PhCs, have been successfully used for label-free refractometric and biosensing applications [1], due to their high-sensitivity to small changes in both dielectric constant and thickness of the materials assembling the PhC structure itself. PhCs featuring different materials (e.g. silicon, polymer, etc.) and architectures (micromirrors, resonant cavities, etc.) have been reported so far for label-free (bio)sensing [1]. Delivery of liquids either over [1, 2] or through [3-5] the PhC has been successfully addressed, the latter envisaging higher sensitivity and lower limit of detection than the former [3]. Pressure-driven operation, through the use of external pumps, has been commonly used for the deliver of liquids in PhCs [2, 4], which limits somehow the applications of PhCs as sensing elements for point-of-care analysis. Nonetheless, capillarity-driven operation, without the use of external pumps, has been also recently demonstrated [5], thus envisaging the development of self-powered drop-and-measure platforms based on PhCs.

Among PhC structures, vertical, high-aspect-ratio silicon/air 1D PhCs able to control light propagation in a plane parallel to the silicon substrate represent an interesting solution for the fabrication of miniaturized platforms to be employed in (bio)sensing [6]. In fact, vertical silicon/air 1D PhCs feature fluidic and optical paths that are inherently independent, the former being through the air voids and the latter perpendicular to the air voids. This enables the integration of both fluidic and optical paths on the same chip as the PhC structure and envisages, in turn, the effective possibility of developing 1D PhC platforms with on-chip fluidic and optical networks. Several cases of 1D PhC platforms making use of vertical 1D PhCs fabricated by dry etching technology, namely deep reactive ion etching, or wet etching technology, namely chemical etching, have been reported for optical applications [7-9]. A few examples have also been reported for sensing applications [10-12].

Recently, deep-etching of complex silicon microstructures and microsystems with sub-micrometer accuracy at aspect-ratio values beyond standard dry and wet etching technologies has been demonstrated by electrochemical micromachining technology (ECM) [13] for both biomedical [14, 15] and photonic [16, 15] applications. ECM technology allows silicon microstructuring to overcome limitations of modern dry etching technologies, with the further advantage of the low cost of ECM technology with respect to dry etching technologies, which could make advanced silicon microstructuring available in any lab.

In this work we review 1D PhC all-silicon platforms, fabricated by electrochemical micromachining technology (ECM) and integrating vertical, silicon/air 1D PhCs together with fluidic and optical paths, for refractometric and biosensing applications both under pressure-driven and capillarity-driven operation [4, 5].

## II. FABRICATION OF ONE-DIMENSIONAL PHOTONIC CRYSTAL PLATFORMS BY ELECTROCHEMICAL MICROMAHINING

A typical 1D PhC all-silicon platform for (bio)sensing applications fabricated by electrochemical micromachining technology integrates a vertical silicon/air 1D PhC operating in the near-infrared region, which is exploited as a label-free sensing element and consists of a spatial repetition (in the micrometer scale) of high-aspect-ratio silicon walls and air-gaps. The 1D PhC is connected to two large reservoirs through a simple fluidic network, which are used to effectively infiltrate the 1D PhC with liquids. The platform also integrates an optical path, perpendicularly to the fluidic path, for either free-space or fiber optics interrogation of the 1D PhC in reflection and transmission.

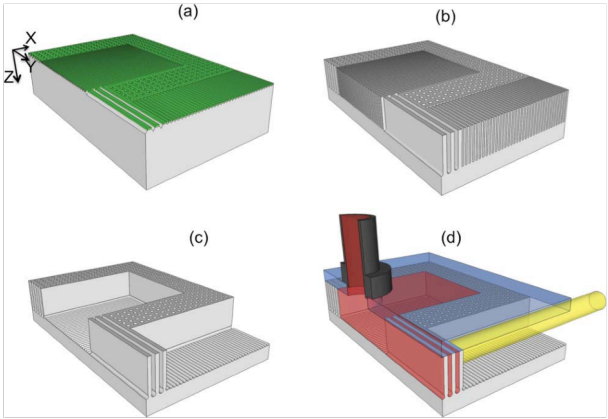

Figure 1. Main technological steps for the fabrication of 1D PhC platforms by ECM technology: (a) pattern definition by standard lithography and seed pattern formation on the silicon surface by KOH etching; (b) deep anisotropic etching of the seed pattern by electrochemical etching and fabrication of high aspect-ratio microstructures; (c) isotropic etching of the bottom of the fabricated high aspect-ratio microstructures by electrochemical etching and release of part of the etched microstructures from the substrate; (d) bonding of a cover either glass or polymer slab, provided with fitting ports. Reproduced from Ref. 4 with permission from The Royal Society of Chemistry.

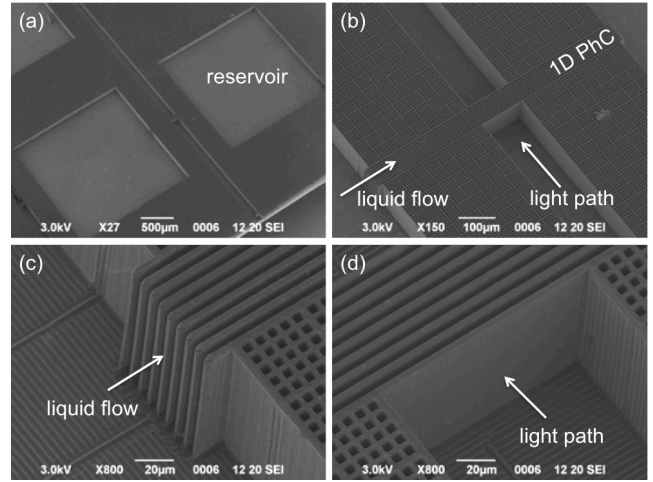

Figure 2. Bird-view SEM images at different magnification of a 1D PhC platform fabricated by ECM technology and integrating a vertical silicon/air PhC with spatial period of 8 μm, height and width of 90 μm and 130 μm, respectively. The 1D PhC directly enters two large reservoirs positioned at both sides and faces two fiber grooves with width of 130 μm suitable for the PhC characterization in reflection and transmission. Reproduced from Ref. 4 with permission from The Royal Society of Chemistry.

The fabrication of 1D PhC platforms by ECM technology is carried out according to the technological process detailed in [4]. A sketch of the main technological steps of the fabrication process is reported in Fig. 1. The starting material is n-type silicon of orientation (100), with a thin (100 nm thick) silicon dioxide layer on top. The layout of the platform to be fabricated is patterned on a photoresist layer by standard lithography, transferred to the silicon dioxide layer by buffered HF (BHF) etching through the photoresist mask, replicated into the silicon surface by potassium hydroxide (KOH) etching through the oxide mask (Fig. 1a), and finally grooved into the silicon bulk by electrochemical etching in aqueous electrolytes containing traces (5% by vol.) of HF (Fig. 1b and 1c). The electrochemical ethcing consists of a single-etching-step with an initial anisotropic phase (Fig. 1b), which is used to etch the seed pattern deep in the substrate and create high aspect- ratio microstructures, and a final isotropic phase (Fig. 1c), which is used to release sacrificial structures from the substrate while leaving functional structures anchored to it. Either glass or polymer slab with suitable inlet/outlet ports aligned to the reservoirs is used as the cover of the platform (Fig. 1d).

Fig. 2 shows scanning electron microscopy (SEM) bird-view images of a 1D PhC platform fabricated by ECM technology, at different magnifications, in which it is possible to appreciate the overall structure of the platform with the 1D PhC in between the two reservoirs, at the intersection of fluidic and optical paths (Fig. 2a). Fig 2b shows a magnification of the 1D PhC structure, which extends up to the reservoirs for this specific platform. Fig. 2 c and 2c provides details of the liquid flow inlet and of the 1D PhC structure, respectively, and allows to clearly appreciate the vertical structure and high-aspect-ratio of the 1D-PhC, as well as the high accuracy in microfabrication and the high quality of the silicon surfaces that are achieved by ECM technology. Noteworthy, as shown in Fig. 3, 1D PhC platforms differing for both fluidic and optical sections can be also fabricated by ECM technology, after a proper layout is defined.

III. PRESSURE-DRIVEN OPERATION OF ONE-DIMENSIONAL PHOTONIC CRYSTAL PLATFORMS

Optical characterization of ECM-fabricated all-silicon 1D PhC platforms under pressure-driven operation is carried out by measuring the reflection spectra of the 1D PhC upon injection of different liquid solutions through the air-gaps. Specifically, ethanol–water mixtures with different concentrations of ethanol are used as the benchmark for refractometry applications; a sandwich assay based on antigen–antibody interactions for the detection of the C-reactive protein (CRP), at a concentration value of 10 mg L$^{-1}$, which is at the boundary level between physiological and pathological conditions, is carried out as the benchmark for biosensing applications. All the liquids are injected in the 1D PhC of the platform through the use of a syringe pump at flow rate of the order of 10 mL min$^{-1}$.

Experimental measurement of the optical power reflected from the 1D PhC in the near-infrared region upon liquid injection is performed using optical fibers positioned into the fiber groove in front of the PhC-itself [4]. For each liquid, measurement of the reflected power spectrum is performed before infiltration of the PhC with the liquid (empty platform), after infiltration of the liquid in the air gaps of the PhC (filled platform), and after removal of the liquid from the PhC air gaps (empty platform). The protocol is repeated several times so as to investigate both reproducibility and accuracy of the 1D PhC platform operation. After acquisition, reflected power spectra are normalized with respect to an ideal reflector. Experimental reflectivity spectra are low-pass filtered in the Fourier Transform domain in order to increase the signal-to-noise ratio and, in turn, to reduce the effect of measurement noise on the reflectivity peak position. The position of the different reflectivity peaks of the 1D PhC is then automatically detected and plotted against the refractive index value of the liquids injected.

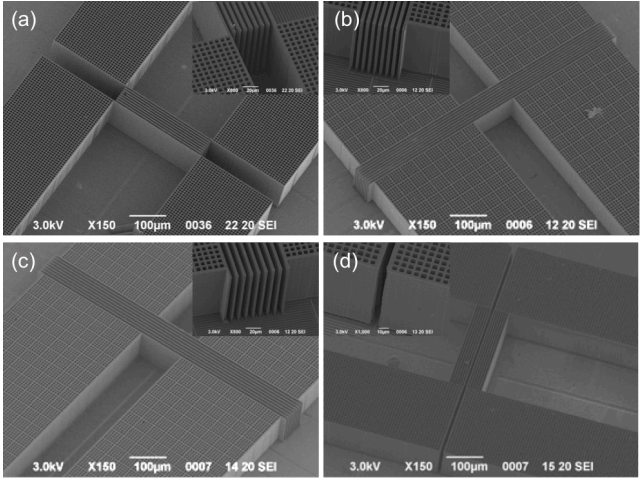

Figure 3. Bird-view SEM images (and inset at higher magnifications) of the core of 1D PhC platforms differing for details concerning both fluidic and optical sections, fabricated by ECM technology. (a) 1D PhC with period of 8 μm, height of 90 μm, and width of 300 μm, connected to the reservoirs through two microchannels with length of 250 μm and width of 60 μm (b) and (c) 1D PhC with periods 8 μm and 10 μm, respectively, and same height and width, 90 μm and 130 μm. In both cases the 1D PhC directly enter the reservoirs. (d) 1D PhC resonant cavity with the half-wavelength air-gap of the cavity (width of 8 μm) connected through two microchannels with length of 250 um to the reservoirs. Reproduced from Ref. 4 with permission from The Royal Society of Chemistry.

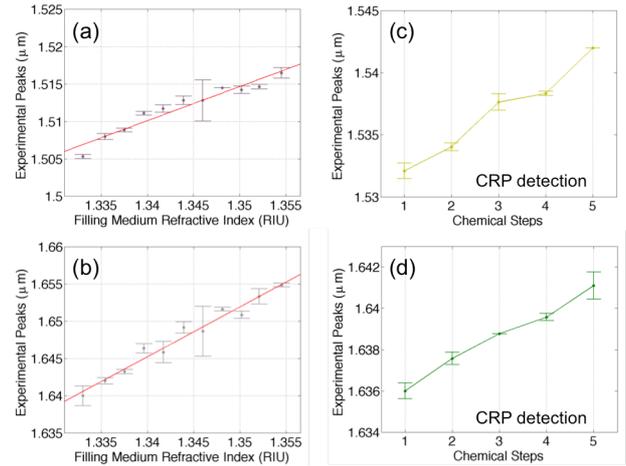

Figure 4. Experimental reflectivity peaks around 1.5 and 1.6 μm of the 1D PhC platform of Fig. 2 under pressure-driven operation: (a, b) as a function of the refractive index value of water/ethanol mixtures injected in the platform; (c, d) after the main chemical steps of the sandwich assay carried out for the detection of the CRP at 10 mg $L^{-1}$, in carbonate buffer before capture-antibody deposition (step1), in HEPES buffer after capture-antibody deposition (step 2), in HEPES buffer after BSA passivation (step 3), in HEPES buffer after antigen incubation (step 4), in HEPES buffer after detection-antibody incubation (step 5). Reproduced from Ref. 4 with permission from The Royal Society of Chemistry.

As to refractometric applications, Fig. 4a and 4b shows experimental reflectivity peaks of the 1D PhC of the platform in Fig. 2 around 1.5 and 1.65 μm (in water), respectively, as a function of the refractive index value of water/ethanol mixtures injected in the platform and filling the air-gaps. An almost linear red shift is observed for the two reflectivity peaks as a function of the increasing refractive index value of the water/ethanol mixtures infiltrating the air gaps. A sensitivity S of 460 and 670 nm/RIU is measured for the two peaks, respectively, associated with a limit of detection of $4.5 \times 10^{-3}$ and $5.8 \times 10^{-3}$ RIU, respectively. As to biosensing applications, Fig. 4c and 4d shows experimental reflectivity peaks of the 1D PhC of the platform in Fig. 2 around 1.53 μm and 1.63 μm (in water), respectively, after the main chemical steps of the sandwich assay carried out for the detection of the CRP at 10 mg $L^{-1}$. Typical standard deviation values around 1 nm are measured over the whole set of peaks, thus indicating that the sandwich assay is fully carried out with good repeatability. A monotonic red shift of the peak position is obtained after each chemical step (steps 1–5), thus well demonstrating that adsorption of biological materials on the PhC surfaces, in general, as well as antigen–antibody interaction at the PhC surfaces, in particular, are successfully monitored through the use of the platform.

IV. CAPILLARITY-DRIVEN OPERATION OF ONE-DIMENSIONAL PHOTONIC CRYSTAL PLATFORMS

Optical characterization of ECM-fabricated all-silicon 1D PhC platforms under capillarity-driven operation is carried out by measuring the reflection spectra of the 1D PhC upon capillary infiltration of different liquid solutions through the air-gaps [5]. Specifically, ethanol–water mixtures with different concentrations of ethanol are used as the benchmark for refractometry applications; Bovine Serum Albumin (BSA) solutions in water at different BSA concentrations are used as the benchmark for biosensing applications. After acquisition, reflected power spectra are normalized with respect to an ideal reflector, then experimental reflectivity spectra are low-pass filtered in the Fourier Transform domain in order to increase the signal-to-noise ratio. The wavelength position of a reflectivity notch around 1.31 μm of the 1D PhC is then automatically detected and plotted against the refractive index value of the liquids injected.

For all the tested mixtures, a calibrated volume of 0.4 μL of liquid is dropped into one of the reservoirs of the platform, which was formerly treated with a Piranaha solution in order modify the chemistry of the silicon surfaces from Si–H to Si–OH, thus switching from hydrophobic to hydrophilic surfaces. The liquid readily infiltrates the 1D PhC structure and the reflectivity spectrum of the 1D PhC is acquired at several sampling times. For each mixture, the procedure is repeated several times such as to infer on both the single-drop and drop-to-drop reproducibility of the measurements.

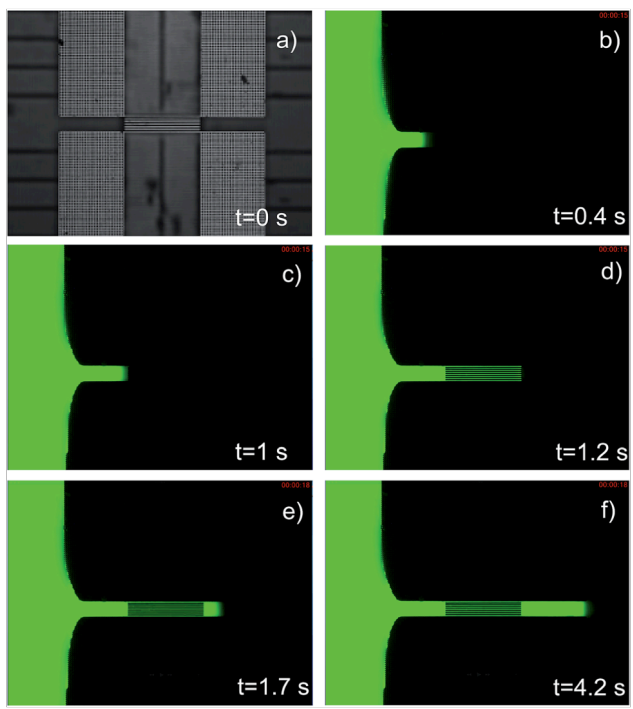

Figure 5. Time-resolved infiltration of a 1D PhC platform fabricated by ECM technology under capillary-driven operation. (a) Bright-field microscopy image and (b–f) fluorescence microscopy images before (a) and after (b–f) dropping 0.4 μL of water containing 0.15 mg mL$^{-1}$ of fluorescein in the left-hand-side reservoir, at different times. Reproduced from Ref. 5 with permission from The Royal Society of Chemistry.

Fig. 5 shows time-resolved infiltration of a ECM-fabricated 1D PhC platform of the type of Fig. 3a under capillary-driven operation, before and after dropping 0.4 μL of water containing 0.15 mg mL$^{-1}$ of fluorescein in the left-hand-side reservoir. The liquid quickly infiltrates the microchannel connecting the reservoir with the 1D PhC, and then the 1D PhC and the second microchannel. Eventually, the liquid stops at the end of the second microchannel, so that the liquid is retained (>30 min for water) in the 1D PhC structure for optical measurements. An average time of 4.2 s is required to fully infiltrate the platform with water.

As to refractometric applications, Fig. 6a shows typical reflectivity spectra around 1.31 μm of the 1D PhC of the platform in Fig. 5 after capillary infiltration with different ethanol–water mixtures. Fig. 6b shows the corresponding calibration curve, which is notch wavelength position versus refractive index of ethanol-water mixtures infiltrated. A red shift of the notch wavelength position is apparent, which can be explained in terms of the increased refractive indices of the mixtures as the ethanol concentration increases. High sensitivity S of 430 nm/RIU and good limit of $1.4 \times 10^{-5}$ RIU are achieved, which are comparable to values obtained for 1D PhC platforms under pressure-driven operation. As to label-free biosensing, Fig. 6c shows typical reflectivity spectra of the 1D PhC of the platform upon capillary infiltration with BSA aqueous solutions at different BSA concentrations. Fig. 6d shows the corresponding calibration curve. A red shift of the notch wavelength position is evident as the BSA concentration increases (from 0.1 to 7%), which can be explained in terms of both the modification of the PhC structure by BSA adsorption on the PhC silicon walls (surface effect) and the increased refractive index of the mixtures flowing through the PhC air-gaps (volume effect). Two different linear regimes are evident in Fig. 6d as the BSA concentration increases from 0.1% to 7%. A higher-sensitivity regime (S of either 3000 nm/RIU or 5.5 nm/%) occurs for lower BSA concentrations (below 1%); a lower-sensitivity regime (S of either 490 nm/RIU or 0.9 nm/%) occurs for higher BSA concentrations (above 1%). The former was attributed to surface effects due to the non-specific adsorption of BSA on the PhC silicon surfaces; the latter was attributed to the refractive index variation of the BSA solutions flowing through the PhC air- gaps. By assuming volume effects constant over the entire range of investigated BSA concentrations, surface sensitivity due to BSA adsorption on the PhC silicon surfaces are estimated to be of 2510 nm/RIU (or 4.6 nm/%) and limit of detection of $2 \times 10^{-6}$ RIU (or 0.001%), which are both comparable to state-of-the-art pressure-driven optical biosensors.

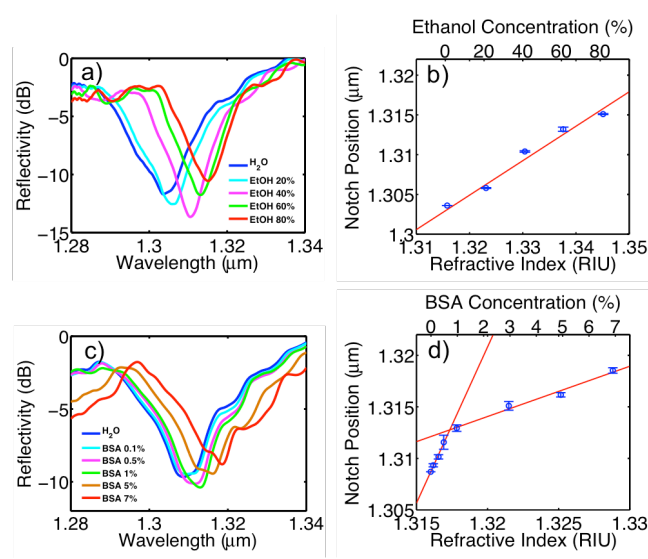

Figure 6. Optical characterization of the 1D PhC platform of Fig. 5 under capillarity-driven operation. (a, b) Typical reflectivity spectra (around 1.3 mm) and calibration curve, respectively, of the 1D PhC upon capillary infiltration with ethanol–water mixtures. (c, d) Typical reflectivity spectra (around 1.3 mm) and calibration curve, respectively, of the 1D PhC upon capillary infiltration with BSA solutions. Reproduced from Ref. 5 with permission from The Royal Society of Chemistry.

## V. Conclusions

In this work 1D PhC platforms fabricated by electrochemical micromachining (ECM) technology for refractometric and biosensing applications were reviewed with reference to both pressure-driven and capillarity-driven applications. In both cases, label-free detection of both surface (anchoring of biomolecules on the PhC surfaces) and volume (variation of the refractive index of the solutions flowing through the PhC air-gaps) effects was successfully demonstrated, with high sensitivity (around $400 - 600$ nm/RIU) and low limit of detection (order of magnitude $10^{-4} - 10^{-5}$ RIU). Point-of-care applications of ECM-fabricated drop-and-measure 1D PhC platforms can be envisaged in the next future.

## Acknowledgment

G. Barillaro would like to thank all the collaborators at the University of Pisa, University of Pavia, and the Institute of Applied Physics "Nello Carrara" of the CNR for their invaluable contribution to the research carried out together on 1D PhC platforms.

## References

[1] Y. Zhao, X. Zhao, and Z. Gu, "Photonic Crystals in Bioassays," Adv. Funct. Mater. 20(18), 2970–2988 (2010).

[2] S. Mandal, J. M. Goddard and D. Erickson, Lab Chip, 2009, 9, 2924.

[3] Y. Guo, H. Li, K. Reddy, H. S. Shelar, V. R. Nittoor and X. Fan, Appl. Phys. Lett., 2011, 98, 041104.

[4] S. Surdo, S. Merlo, F. Carpignano, L. M. Strambini, C. Trono, "Optofluidic microsystems with integrated vertical one-dimensional photonic crystals for chemical analysis," A. Giannetti, F. Baldini and G. Barillaro, Lab Chip, 2012, 12, 4403.

[5] S.Surdo,F.Carpignano,L.M.Strambini,S.Merlo,andG.Barillaro,"Capillarity-driven (self-powered) one-dimensional photonic crystals for refractometry and (bio)sensing applications," RSC Adv. 4(94), 51935–51941 (2014).

[6] S.Surdo,F.Carpignano,G.Silva,S.Merlo,andG.Barillaro,"An all-silicon optical platform based on linear array of vertical high-aspect-ratio silicon/air photonic crystals," Appl. Phys. Lett. 103(17), 171103 (2013).

[7] M.RenilkumarandP.Nair,"Low-loss optical channel drop filters based on high-contrast Si–air photonic crystals by wet anisotropic etching," Appl. Opt. 50(25), E59–E64 (2011).

[8] A.Lipson, E.M.Yeatman,"A 1-D Photonic Band Gap Tunable Optical Filter in (110) Silicon," J. Microelectromech. Syst. 16(3), 521–527 (2007).

[9] J. M. Masson, R. St-Gelais, A. Poulin, and Y.-A. Peter, "Tunable Fiber Laser Using a MEMS-Based In Plane Fabry-Pèrot Filter," IEEE J. Quantum Electron. 46(9), 1313–1319 (2010).

[10] P.S.Nunes,N.A.Mortensen,J.P.Kutter and K.B.Mogensen,"Photonic crystal resonator integrated in a microfluidic system", Opt. Lett., 2008, 33(14), 1623.

[11] R. St-Gelais, J. Masson, and Y.-A. Peter, "All-silicon integrated Fabry–Pérot cavity for volume refractive index measurement in microfluidic systems," Appl. Phys. Lett. 94(24), 243905 (2009).

[12] G. Barillaro, S. Merlo, S. Surdo, L. M. Strambini, F. Carpignano, Integrated Optofluidic Microsystem based on Vertical High-Order One-Dimensional Silicon Photonic Crystals, Microfluidics Nanofluidics, 12, 545-552 (2011).

[13] M. Bassu, S. Surdo, L. M. Strambini, G. Barillaro, Electrochemical micromachining as an enabling technology for advanced silicon microstructuring, Advanced Functional Materials, 22 (6), 1222-1228 (2012).Bbb

[14] L. M. Strambini, A. Longo, A. Diligenti, G. Barillaro, A Minimally Invasive Microchip for Transdermal Injection/Sampling Applications, Lab Chip, 12, 3370-3379 (2012).Bbb

[15] L.M. Strambini, A. Longo, S. Scarano, T. Prescimone, I. Palchetti, M. Minunni, D. Giannessi, G. Barillaro, Self-Powered Microneedle-Based Biosensors for Pain-Free High-Accuracy Measurement of Glycaemia in Interstitial Fluid, Biosensors and Bioelectronics (2014).

[16] Ou Weiying, L. Zhao, S. Surdo, H. Diao, H. Li, C. Zhou, W. Wang, G. Barillaro, 3-D Solar Cells Based on Radial Silicon Heterojunctions Exploiting Microhole Lattices, IEEE Photonics Technology Letters, 25, 19 (2013).

[17] S. Merlo, F. Carpignano, G. Silva, F. Aredia, A. I. Scovassi, G. Mazzini, S. Surdo, G. Barillaro, Label-free optical detection of cells grown in 3D silicon microstructures, Lab Chip, 13, 3284 (2013).